\documentclass[12pt,preprint]{aastex}
                                                                                
\shorttitle{Slow VLBI knots from fast jets}
\shortauthors{Gopal-Krishna, Dhurde \& Wiita}
                                                                                
\begin{document}
                                                                                
                                                                                
\title{Do the mildly superluminal VLBI knots exclude
ultrarelativistic blazar jets?} 
                                                                               
                                                                                
\author{Gopal-Krishna$^{1}$, Samir Dhurde$^{2}$, Paul J.\ Wiita$^{3}$}
\affil{$^1$National Centre for Radio Astrophysics, TIFR, Pune
                University Campus,
Post Bag No.\ 3,   Pune 411007, India; Email: krishna@ncra.tifr.res.in\\ 
$^2$Department of Physics,  University of Pune, Pune 411007, India; Email:
samiris@ncra.tifr.res.in\\
$^3$Department of Physics \& Astronomy, P.O.\ Box 4106,
Georgia State University, Atlanta, GA 30302-4106; Email: wiita@chara.gsu.edu}


                                                                                
                                                                                
\begin{abstract}
We compute the effective values of apparent transverse velocity and 
flux boosting factors for the VLBI radio knots of blazar jets,
by integrating over the angular distributions of these quantities
across the widths of jets with finite opening angles
but constant velocities. For high bulk Lorentz factors ($\Gamma > 10$)
variations across the jet can be quite large if the opening
angle, $\omega$, is even a few degrees on sub-parsec scales.
The resulting apparent speeds are often much lower than those
obtained from the usual analyses that ignore the finite
jet opening angles.  We can thus reconcile the 
usually observed subluminal or mildly superluminal speeds
with the very high  ($\gtrsim 20$) $\Gamma$ factors, 
 required by the inverse Compton
origin and rapid variability of TeV fluxes, 
as well as by intraday radio variability. 
Thus it is possible to associate the VLBI radio knots directly
with shocks in the ultra-relativistic main jet flow, 
without invoking very rapid jet deceleration on parsec scales, 
or extremely unlikely viewing angles.

\end{abstract}
                                                                                

\section{Introduction}

In order to avoid excessive photon--photon collisions, 
the highly variable TeV emission from blazars has been interpreted in terms of
inverse Compton radiation emerging from ultrarelativistic jets
with bulk Lorentz factors, $\Gamma \sim 15-100$ (e.g., Krawczynski,
Coppi \& Aharonian 2002; Piner \& Edwards 2004;
Ghisellini, Tavecchio \& Chiaberge 2004).
Earlier, very high  values of $\Gamma$ ($\sim 100$) 
were inferred from the intraday
radio variability (IDV) which is a common feature of blazars (Wagner \& Witzel
1995; Kraus et al.\ 2003).  It now appears that the bulk of this IDV may arise
from interstellar scintillations (e.g., 
Kedziora-Chudczer et al.\ 2001; 
 Bignall et al.\ 2003). 
Nonetheless, the required microarcsecond angular 
sizes of the scintillating components would still need $\Gamma > 30$,
and probably substantially larger,
in order to reconcile apparent 
brightness temperatures, $T_B \sim 10^{13-14}$K (Blandford 2002;
Rickett et al.\ 2002)
with the canonical
limit of $T_B < 10^{12}$K needed to avoid the inverse Compton catastrophe
(e.g., Kellermann \& Pauliny-Toth 1969). Direct evidence for
high values of $T_B$ comes from the VSOP survey (Horiuchi et al.\ 2004). 
Standard models for gamma-ray bursts also invoke bulk ultrarelativistic
jet flows with $\Gamma \sim 100-1000$ (e.g., 
Sari, Piran \& Halpern 1999; M{\'e}sz{\'a}ros 2002).

In contrast, the only direct probe of extragalactic jet motion, 
namely the radio knots detected by Very Long Baseline Interferometry (VLBI),
reveal typical proper motions corresponding to apparent 
speeds,
$v_{app} \equiv c\beta_{app}$, of 5--10$c$ or less 
(Jorstad et al.\ 2001; Cohen et al.\ 2003; Giovannini 2003). 
Recent measurements have shown that roughly 
one-third to one-half of the VLBI components
measured in TeV blazars are found to be subluminal or even 
essentially stationary (Piner \& Edwards 2004; Giroletti et al.\ 2004).  
As emphasized by Piner \& Edwards (2004),
the apparent lack of relativistically moving shocks that are assumed to be
responsible for the high energy flaring activity, and are convincingly
seen in most EGRET blazars (Kellermann et al.\ 2004), 
is intriguing.

One possible explanation for the slow VLBI components of
the TeV blazars would be a dramatic deceleration of the jet
between sub-pc and pc-scales (e.g., Georganopoulos \&
Kazanas 2003); but then the problem
becomes to understand how such a deceleration is avoided in the
case of EGRET blazars for which distinctly higher apparent speeds are
measured (Piner \& Edwards 2004; Kellermann et al.\ 2004).
Essentially perfect alignment (to within $1^{\circ}$) of the TeV blazar jets 
in our direction is another possible way out; however, given
their substantial number, angles of a few degrees are much more likely
(Piner \& Edwards 2004).
Other attempts to explain the relatively
slow apparent motions of the knots involve postulating a spine and sheath
geometry for the jets as was proposed for FR I sources 
(e.g., Sol,  Pelletier \& Asseo 1989) 
such that the radio knots are frequently
associated with the slower moving outer layer 
(Komissarov 1990;  Laing et al.\ 1999; Chiaberge et al.\ 2000; 
Ghisellini et al.\ 2004).  
Some support for
this picture comes from the limb-brightening marginally observed in a
few parsec-scale jets (e.g., Giovannini 2003; Giroletti et al.\ 2004).

Here we shall argue that the 
observed mildly superluminal, or even subluminal, speeds
of the parsec-scale VLBI knots do not necessarily demand 
that the knots be associated exclusively with an outer 
sheath; the knots may well be formed by shocks occurring
in the energetic,
ultra-relativistic flow of the spine.  
Nor is any overall slowing, or an
extraordinarily close alignment, required to account for
the observed substantial
fraction of slow VLBI knots.
Since the jets on the parsec scale are likely to be a few degrees wide,
if they are pointed close to our direction and have a high bulk Lorentz factor, one expects very large
variations of the apparent velocity and Doppler boosting across the
face of the jet or blob.
The calculations we present here show
that when the distribution of apparent velocity across the jet
cross-section is weighted by the corresponding
Doppler boosting factor distribution, as is
appropriate when the jet cannot be transversely resolved,
there can be a marked overall reduction in 
the resulting apparent velocity, compared
to the usually  assumed case where both the Doppler boosting factor 
 and the apparent velocity are taken to be
constant across the jet.  Therefore, the Lorentz factors,
$\Gamma \gtrsim \beta_{app}$,
usually inferred from the VLBI motion data, often may be gross underestimates
of the actual bulk (or pattern) Lorentz factors.

\section{The Model}

We now compute the boosting-weighted apparent velocity $v_{app,w}$ and 
observed flux density, $S_{o,w}$.  
The standard Doppler boosting factor 
for a jet is
$\delta = [\Gamma(1-\beta {\rm cos}\theta)]^{-1}$,
where $\beta \equiv v/c$, $\theta$ is the angle between the jet's axis
and the observer's  line-of-sight, and
$\Gamma \equiv (1-\beta^2)^{-1/2}$ (e.g., Scheuer \& Readhead 1979).
The observed flux, $S_{\nu,o}$, is related to the emitted flux,
$S_{\nu,e}$ via
$S_{\nu,o} = \delta^{n-\alpha} S_{\nu,e}$,
where $n = 2$ for a
continuous jet but $n = 3$ for a discrete ``plasmoid'' or shocked
emitting region
(as the VLBI components are usually treated), and $\alpha$ is the
spectral index, defined by $S_{\nu} \propto \nu^{+\alpha}$.
For simplicity, we will assume $\alpha = 0$
for the VLBI knots, and will ignore
the cosmological effects implemented through factors of
$(1+z)$. We will further
ignore any possible distinction between
a bulk velocity and a pattern speed; none of these simplifications
affects our qualitative argument.  

The apparent transverse velocity of a knot in
the jet is 

\begin{equation}
v_{app} = \frac{v~ {\rm sin}~\theta}{1 - \beta~ {\rm cos}~\theta},
\end{equation}

\noindent and the maximum value of $v^{max}_{app} = \Gamma c$ when $\theta = 1/\Gamma$.
Since the typical values of $v_{app}$ are under 10$c$, the usual
statistical studies of VLBI component velocities seem to imply that
the dominant bulk (or pattern) Lorentz factors are also $< 10$ (e.g.,
Vermeulen \& Cohen 1994; Kellermann et al.\ 2004). 

But if the jet has a finite opening angle, $\omega$, corresponding
to a solid angle, $\Omega$, as opposed to an 
infinitesimal opening angle, then each small element of the jet cross-section
is boosted
by a different amount because of having a different misalignment from the
line-of-sight, even if all elements have exactly the same bulk velocity,
as we shall assume here. 
Hence, we must integrate over the solid angle of the jet to
obtain the boosting-weighted values of the key observed quantities.
The weighted flux is 

\begin{equation}
S_{o,w} = \int_{\Omega} \delta^n(\Omega^{\prime}) ~S_{e}(\Omega^{\prime}) ~ d\Omega^{\prime}~ \equiv~ A(\theta) S_{e},
\end{equation}

\noindent where we have suppressed the subscript $\nu$, and in the second equality
explicitly taken $S_{e}$ to be independent of $\Omega^{\prime}$,
 and defined the
mean amplification factor, $A(\theta)$.  

We then perform an integration of the (boosted) flux weighted
apparent velocity over the jet cross-section to obtain the
weighted observed value of the apparent velocity of the jet,

\begin{equation}
{\vec \beta}_{app,w} = 
{\frac{1}{S_{o,w}}} \int_{\Omega} ~{\vec \beta}(\Omega^{\prime})~
 \delta^n(\Omega^{\prime})~ S_{e} ~  d\Omega^{\prime} .  
\end{equation}

\noindent Note that the resultant vector is along the line joining the
directions of the blazar nucleus and the center of the jet's cross-section.
In Fig.\ 1 we show the dependence of $\beta_{app,w}$ on $\theta$ for
$\Gamma = 10, 50$, and $100$, taking $n = 3$.
                                                                                

For any combination of $\Gamma$, $\omega$, and $n$ we can now compute the
probability distribution of $\beta_{app,w}$. For this an 
integration needs to be done over strips of solid angle represented
by concentric annuli with radii $\theta$ around the source direction.
There will be an unique value of $\beta_{app,w}$ for each such
annulus, which is given by the magnitude of Eq.\ (3).  This value will be
weighted by the number of sources for that annulus that would be visible
in a flux-limited sample.  This number, or when normalized,
this probability, $p(\theta; \Gamma, \omega, n)$, will be proportional to both
the solid angle of the annulus and the enhancement of
the source counts due to Doppler boosting, $A$, averaged over the
jet's cross-section, when the latter is centered on that annulus.
We ignore any contribution from a counter-jet, which should be extremely
de-boosted. The enhanced surface density of these core-dominated
sources, which are detected in  high-frequency (cm-wavelengths) flux-limited
surveys, is proportional to $A^q$, where $q$ is the exponent 
in the integrated counts of radio sources (e.g., Cohen 1989),
$N(S_e) dS_e \propto S_e^{-q} dS_e$;
we take $q =3/2$, which is strictly
applicable for Euclidean space, but is a good approximation for these
core-dominated radio source samples (e.g., Fomalont et al.\ 1991).
The number of sources seen in a flux-limited sample for a
particular misalignment angle $\theta$ will thus be
related to the solid angle subtended by that annulus,
multiplied by $A^q$.  We find the probability to be

\begin{equation}
p(\theta) d\theta \propto {\rm sin}~\theta~ A^q(\theta) d\theta.
\end{equation}

\noindent Eq.\ (4) can also be used to provide the distribution of
$\beta_{app,w}$ since there is a unique value of it 
for every $\theta$.

Our key result is shown in Fig.\ 1, where the probability of
observing any $\beta_{app}$ greater than a given value is shown;
this is obtained by integrating Eq.\ (4) and subtracting the
normalized cumulative probability distribution function from unity.
Distributions for the canonical case of a narrow cylindrical jet, $\omega = 0$, 
are plotted for comparison. 

The results for $n=2$, which are applicable to the continuous jet
case, are quite similar: the locations of
the peaks in the $\beta_{app,w}$ curves 
are hardly shifted, but the maximum values at those peaks are
somewhat lower because of the weaker boosting.  
The cumulative probabilities
decline slightly less rapidly
than those plotted in Fig.\ 1 for low $\beta$, 
but more rapidly at high $\beta$ (because of the lower
peak values) thus yielding fewer very
low values of $\beta_{app,w}$ for  given  
$\Gamma$ and $\omega$.  

\section{Discussion and Conclusions}

We have presented  computations
indicating that a significant reduction in the apparent motion
can be expected for VLBI radio knots associated with an
ultrarelativistic jet by taking into account the jet's
opening angle.  On the pc-scale which could be 
identified with the collimation regime, the jets are likely to
be at least several degrees wide; they may be substantially wider,
as indicated for the best resolved case of M87 
(Junor, Biretta \& Livio 1999).

For this situation, we find that
the apparent transverse velocity peaks at 
lower values, and these peaks can occur at significantly
greater  angles to the line of sight than they do
 for the usually assumed case of an infinitesimally small
jet opening angle on parsec scales (Fig.\ 1).  
These trends become stronger 
for high $\Gamma$, for then 
$v_{app,w}$ is sharply peaked around some $\theta > 1/\Gamma$.
A reversal of the sign of contributions
to $v_{app,w}$ arising from some parts of the jet's
cross-section occurs if $\theta < \omega/2$.  
The resulting cancellation,
which is further enhanced because of the sharply declining
Doppler boost with $\theta$, can often lead to a fairly
drastic reduction in the apparent velocity of the knots,
as compared to the canonical peak value of $\beta_{max} \simeq \Gamma$
(Fig.\ 1).

In the core dominated samples that are characteristic
of BL Lacs, the usual expectation, based on $\omega = 0$
jets, is to find sources with $v_{app}$ widely
distributed up to $\Gamma c$, but with
the distribution actually skewed toward the higher 
part of that range (as
shown in Fig.\ 1; see also Vermeulen \& Cohen 1994).  
Thus, when the $v_{app}$'s measured
for VLBI knots are frequently
$< 10$c,  one of the following conclusions
discussed in $\S$1 is usually drawn:  bulk $\Gamma$ values 
are modest; pattern velocities due to shock
motions are  slower than the jet flow; the VLBI knots are
associated with a slower sheath of the jet;
the angle to the line of sight is extremely small.  
While all these are possible, here we have shown that 
none is necessary; rather,
if the jet has a modest full-opening angle ($\omega$) on the scales
probed by VLBI, then there is a very large
reduction in the probability of measuring apparent
velocities approaching $\Gamma c$.  For instance, 
even for the extreme case of
$\Gamma = 100$ and a modest jet opening angle,
$\omega = 5^{\circ}$, over 73\%
of the radio components would be detected with $v_{app} < 10c$,
while for $\omega = 10^{\circ}$, over 87\% would fall
into this category. Over 41\% (for $\omega = 5^{\circ}$)
and over 69\% (for $\omega = 10^{\circ}$) would 
actually be seen as subluminal sources.  
Similarly, for  $\Gamma = 50$
and $\omega = 5^{\circ}$, over 64\%
of the sources would be detected with $v_{app} < 10c$,
the median value is $\beta_{app} = 6$, and still some 15\% 
would appear as subluminal sources.

Therefore, the predominance of marginally superluminal or
even subluminal VLBI knots among 
 TeV blazars does not imply
that these radio knots cannot be physically associated 
with their ultrarelativistic jets.
Instead, a combination of  high  $\Gamma$ factors and 
modest jet opening angles  can just as well
explain the preponderance of low $v_{app}$ values.
At the same time, high $\Gamma$ factor jets ($> 15$)
are needed by the standard one-zone models to  efficiently produce
the TeV photons by inverse Compton scattering (such relatively
modest values emerge from models only when de-reddening of the TeV spectrum
by the IR background is ignored) and higher values
($\Gamma > 40$) are usually required when the TeV spectrum is 
appropriately de-reddened (e.g.,
Krawczynski et al.\ 2002; Ghisellini et al.\ 2004).
Only multi-zone (rapidly decelerating or spine-sheath) models 
can give satisfactory fits to TeV blazar
spectra with $\Gamma \sim 15$ even with de-reddening 
(Georganopoulos \& Kazanas 2003; Ghisellini et al.\ 2004).

Clearly, the applicability of our interpretation is
not restricted to TeV blazars.  Relatively slow proper
motions ($v < 3c$) are now known to occur frequently
among the radio selected samples of quasars (for instance,
the 2-cm VLBA survey, Kellermann et al.\ 2004).
In our picture, such slow motions would not be at variance
with the values of $\Gamma > 30$ which seem to
be needed to explain the scintillating radio components
of IDV blazars (e.g., Blandford 2002).
The values of $\theta$ for all blazars should be small, but are
more likely to be a few degrees rather than less than 1 degree, as shown
by the Monte Carlo simulations of Lister \& Marscher (1997) and
by the detection of quiescent x-ray emission (Giebels et al.\ 2002),
so explaining the  slow apparent motions through extremely precise
jet alignment is a less attractive alternative.            

We recall that essentially all models for gamma-ray bursts
also invoke very high $\Gamma$ factors for their jets (e.g.,
M{\'e}sz{\'a}ros 2002). If the processes that accelerate jets on these very
different scales are basically similar, then this observation from
GRBs may provide further support for the high $\Gamma$ scenario
in quasars and blazars (e.g., Kundt \& Gopal-Krishna 2004).

Although our picture explains the preponderance of slow
velocities, there is always a tail to the computed $v_{app}$
distribution which
extends up to a substantial fraction of $\Gamma c$.
While the highest well measured values of $v_{app}$ do
reach up to 35--40$c$ (Cohen et al.\ 2003; Kellermann et al.\ 2004), there is 
a strong possibility of bias against
picking out VLBI components showing anomalously high
apparent velocities.  The difficulty in tracking such
fast moving knots in the usually sparsely temporally
sampled databases
is compounded by their expected rapid fading.

Upper limits on $\omega$ for powerful blazars 
can be set, since
the total jet power, $L_j$, is related to the {\it inferred} bolometric
luminosity $L_{bol}$ via $L_j = (L_{bol}/\epsilon)(\Omega/4 \pi) =
L_{bol}\omega^2/\epsilon$,
with $\epsilon$ the efficiency of converting jet power into
radiation.  Since $\epsilon < 0.1$ is expected  and $L_{j}$
should not exceed the Eddington limit, we have
$\omega < 0.37[(\epsilon/0.1)(M_{BH}/10^9M_{\odot})/(L_{bol}/10^{48}{\rm erg~ s}^{-1})]^{1/2};$
here $\omega$ is in radians and we have scaled $L_{bol}$ and
the supermassive black hole mass, $M_{BH}$, by typical values. 
Unless $\epsilon < 0.01$, $\omega > 5^{\circ}$ should be allowed
for even very powerful blazars.


It is interesting to examine our picture in the context of the
``spine + sheath'' model of the jet, e.g., as proposed by 
Chiaberge et al.\ (2000) to bring the  data into accord with
the orientation based unification model (e.g., Gopal-Krishna 1995; 
Urry \& Padovani 1995). By considering 
the nuclear emission in radio, optical and x-ray bands from BL 
Lacs and their presumed misaligned counterparts, the FR I radio galaxies,
they inferred that a mildly relativistic ($\Gamma = 1 - 2$) sheath 
component is the prime contributor to the observed nuclear emission 
from FR I RGs. At the same time they showed that the 
spectral energy distribution
of the beamed counterparts, i.e., BL Lacs, all the way from
radio to $\gamma$-rays can be modeled in terms of a relativistic
spine component moving with a bulk-Lorenz factor $\Gamma \simeq 15-20$. 
Still higher $\Gamma$ values have been estimated in several other 
studies, particularly for TeV blazars, as mentioned above. In this 
work we have only discussed the spine component and
tried to address the question: can such large $\Gamma$ 
values be reconciled with the statistics of superluminal motion of the 
VLBI knots of blazars/BL Lacs, which suggest a typical $\Gamma \sim 3$ (e.g., 
Piner \& Edwards 2004; Kellermann et al.\ 2004)? 
We have argued that the discrepancy can be resolved
by considering a modest opening angle ($\sim 5^{\circ}$) for the jet/spine
on parsec scales. The blazar observations can then be understood 
without invoking a rapid jet deceleration on such small scales, or 
relegating the VLBI knots exclusively to a slower sheath (and thus 
totally decoupling them from the shocks occurring within the spine), 
or postulating extremely unlikely viewing angles for the jet. 
 
The main question addressed here can be rephrased
as follows: if indeed the fast spine component alone is relevant for 
the observed blazar emission, then how large a typical Lorentz factor
can be reconciled with their VLBI data? We have argued that even when 
the VLBI knots are associated with shocks in the spine itself, their 
usually observed modest speeds ($v_{app} \sim 3-5c$) would not be
inconsistent with $\Gamma > 30 - 50$ of the spine, provided one 
takes into account a $5 - 10^{\circ}$ opening angle of the spine
on the pc-scale. 
Thus, the VLBI results can be reconciled with the 
TeV and IDV observations indicating such large $\Gamma$ factors (\S 1).
                                    
One prediction of this picture is that
 when adequate resolution and sensitivity is achieved 
so as to be able to 
transversely resolve the fastest moving VLBI scale knots, 
then different portions of those knots would sometimes evince  different
apparent velocities because variations across the jet opening
angle could then be detected.  If sufficient dynamic range 
also becomes available, such components might be seen to fragment,
or, perhaps more likely, appear to be smeared out quickly.



\acknowledgements 
SD is grateful to NCRA for a Project Studentship.
PJW appreciates the hospitality provided at NCRA and at Princeton
University. 
PJW's efforts were partially supported by RPE
funds at GSU.




                                                                                
 

\newpage
\begin{figure}
\plotone{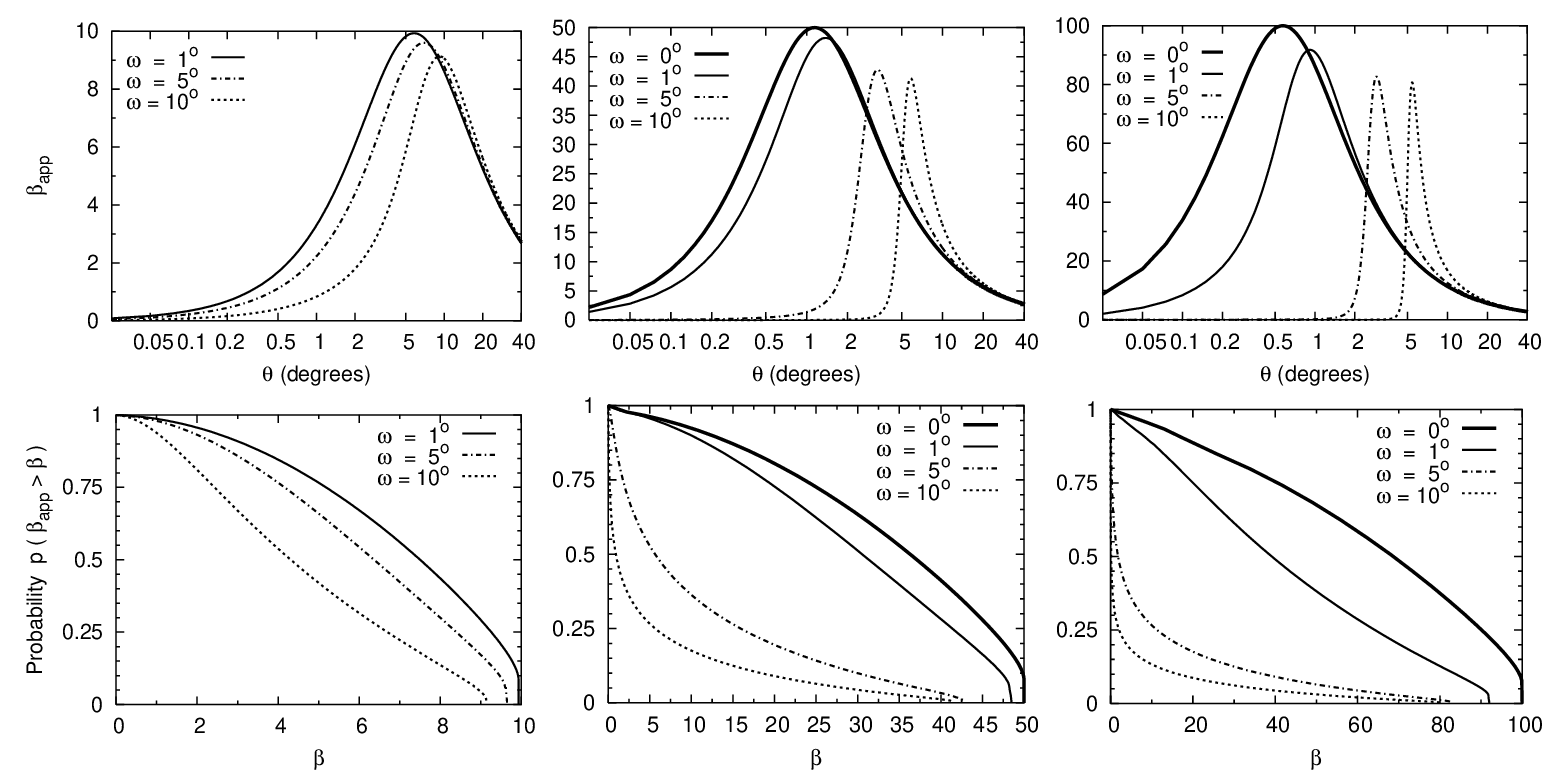}
\caption{Upper panels: distributions of $\beta_{app,w}$ 
against $\theta$ for $\Gamma =$ 10
(left), 50 (center) and 100 (right).  Results for jet opening angles,
$\omega = 0, 1, 5$ and $10$ degrees are shown.   Lower panels: 
cumulative probability for  $\beta_{app,w} > \beta$ for the same values
of $\Gamma$ and $\omega$.  In the left panels the
results for $\omega = 0^{\circ}$ and $\omega = 1^{\circ}$ are 
indistinguishable, so only the latter are labeled.
}
\end{figure}
\end{document}